\documentclass[preprint,aps,amsmath,amssymb,showpacs,pre,superscriptaddress]
              {revtex4} 

\usepackage{graphicx}
\begin{document}
\title{Grand-canonical and canonical solution of self-avoiding walks
  with up to 3 monomers per site on the Bethe lattice} 
\author{Tiago J. Oliveira}
\email{tiagojo@if.uff.br}
\affiliation{Instituto de F\'{\i}sica
}
\author{J\"urgen F. Stilck}
\email{jstilck@if.uff.br}
\affiliation{Instituto de F\'{\i}sica and
National Institute of Science and 
Technology for Complex Systems
\\
Universidade Federal Fluminense\\
Av. Litor\^anea s/n\\
24210-346 - Niter\'oi, RJ\\
Brazil}
\author{Pablo Serra}
\email{serra@famaf.unc.edu.ar}
\affiliation{Facultad de Matem\'atica, Astronom\'{\i}a y F\'{\i}sica\\
Universidad Nacional de C\'ordoba and IFFAMAF  - CONICET\\
C\'ordoba - RA5000\\
Argentina}
\date{\today}

\begin{abstract}
We solve a model of polymers represented by self-avoiding walks on a
lattice which may visit the same site up  to three times in the
grand-canonical formalism on the Bethe lattice. This may be a model
for the collapse transition of polymers where only interactions
between monomers at the same site are considered. The phase diagram of
the model is very rich, displaying coexistence and critical surfaces,
critical, critical endpoint and tricritical lines, as well as a
multicritical point. From the grand-canonical results, we present an
argument to obtain the properties of the model in the canonical
ensemble, and compare our results with simulations in the
literature. We do actually find extended and collapsed phases, but the
transition between them, composed by a line of critical endpoints and
a line of tricritical points, separated by the multicritical point, is
always continuous. This result is at variance with the simulations for
the model, which suggest that part of the line should be a
discontinuous transition. Finally, we discuss the connection of the
present model with the standard model for the collapse of polymers
(self-avoiding self-attracting walks), where the transition between
the extended and collapsed phases is a tricritical point.
\end{abstract}

\pacs{05.40.Fb,05.70.Fh,61.41.+e}

\maketitle

\section{Introduction}
\label{intro}

Polymers may be modelled as self-avoiding walks on a lattice. Each
walk visits a sequence of first-neighbor lattice sites and may be
viewed as a chain of monomers located on the lattice sites linked by
bonds on the lattice edges. This model was already considered by Flory
in his pioneering work on polymers \cite{f66}, and later De Gennes
discovered a mapping between this model and the $n$-vector model of
magnetism in the formal limit $n \to 0$, which allowed him to apply
the renormalization group formalism to this problem \cite{dg72}. The
ideas of scaling and universality are very important in this field
\cite{dg79}.

If the polymer is placed in a poor solvent, the interactions between
the molecules (monomer-monomer, monomer-solvent, and solvent-solvent)
penalize energetically the monomer-solvent contacts. In a lattice
model, this may be
studied considering an {\em effective} attractive interactions between
monomers located on first-neighbor sites of the lattice which are not
consecutive along a chain. These interactions compete with the
repulsive excluded volume interactions (which lead to the
self-avoidance constraint), and at sufficiently low temperatures the
chain may undergo a collapse transition, from an extended to a more
compact configuration. For example, the exponent $\nu$ which describes
the behavior of the mean square end-to-end distance of the chain as a
function of the molecular weight (number of monomers) $M$, $\langle
R^2 \rangle \approx M^{2\nu}$, changes from a larger value in the
extended configuration to $1/d$, where $d$ is the dimensionality of
the lattice, in the collapsed state. The temperature at which this
transition happens is called the $\theta$-temperature \cite{f66}. 
It turns out to be particularly interesting
to consider this model of self-attracting self-avoiding walks
(SASAW's) in the grand-canonical ensemble \cite{dg79}, since the
polymerization transition in the chemical potential $\times$
temperature phase diagram is found to change from continuous, at high
temperatures to discontinuous below the $\theta$-temperature. This is
consistent with the description of the behavior of the exponent $\nu$
above, since at the transition the density of monomers vanishes for
$T>T_{\theta}$ and is finite for $T<T_{\theta}$. Thus, the
$\theta$-point may be recognized as a tricritical point. Much is known
about this tricritical point in two dimensions \cite{ds85} and its
exact tricritical exponents were found through the study of a diluted
polymerization model \cite{ds87}. On the square lattice, for a model
where the attractive interactions are between {\em bonds} of the chain
on opposite
edges of elementary squares, there are indications that an even
richer phase diagram is found, with an additional dense polymerized
phase \cite{sms96,p02}.

Some time ago, a model was introduced to study interacting polymers
where only {\em one-site} interactions are present  \cite{kpor06}, in
opposition to the usual SASAW's model, where we have interactions between
monomers at first-neighbor sites. Usually lattice models for fluids
may be viewed as a consequence of the partition of space into small
cells (cellular model), of molecular size. In the SASAW's model, such a
cell may be occupied by a single monomer or empty. Now if the cells
are larger, they may be occupied by more than one monomer (we will
call this the {\em multiple monomers per site} (MMS) model) , and the
interactions may be supposed to occur only between monomers in the same
cell. If the bonds are larger than the size of the cells, two monomers
in the same cell may not be connected by a bond. Thus, in the new
model each lattice site may be either empty or 
occupied by $1,2,\ldots,K$ monomers, and an attractive interaction
exists between each pair of monomers on the same site. This model may
be viewed as a generalization of the Domb-Joyce model, where also
multiple monomers may be placed on the same lattice site
\cite{dj72}. In \cite{kpor06} two versions of the model were studied
using canonical simulations: in the RA (immediate reversals allowed)
model there 
are no additional restrictions on the walks, but in the RF (immediate
reversals 
forbidden) model configurations where the walk leaves one site,
reaches a first neighbor and returns to the original site are not
allowed. The simulation lead to particularly interesting results for
the RF model on the cubic lattice, with two distinct collapse
transitions present in the phase diagram. The precise nature of these
transitions, as well as the nature of the multicritical point where
the two transition lines meet, could not be found, although one of the
collapse transitions seems to be discontinuous. 

Recently, the RF and
RA models in the grand-canonical ensemble were solved on the Bethe
lattice \cite{ss07} for the case $K=2$. The parameter space for this
model is defined by the statistical weights $\omega_i$, $i=1,2$, of
sites occupied by $i$ monomers (the weight of empty sites is equal to
one). In the solution of the RF
model  the continuous polymerization transition at high temperatures
ended at a tricritical point, similar to what is observed for the
SASAW's model. An additional polymerized phase (DO) appears at higher values
of the statistical weight of double occupied sites $\omega_2$, where
only empty 
and double-occupied sites are present. Later, the RF model, still for
$K=2$, was solved on the Husimi lattice \cite{oss08}. It is
expected that this solution should be closer to the thermodynamic
behavior observed on
regular lattices. The phase diagram found is similar to the one
obtained in the Bethe lattice solution, although the region of
stability of the DO phase is smaller. It is also worth to mention that
a rather unphysical result found on the Bethe lattice, where the
tricritical collapse transition point is found for {\em vanishing}
interactions between monomers, is corrected in the Husimi lattice,
where this point is located in the region of {\em attractive}
interactions, as expected.

In this paper we solve the model for $K=3$ on the Bethe lattice. This
generalization of the previous work allows us to compare our results
with the simulations described by Krawczyk et al
\cite{kpor06}. Actually, the correspondence between the canonical
ensemble, in which the simulations were done, and the grand-canonical
ensemble used in this paper is not straightforward for the polymer
models, and we will discuss this matter in detail below. Basically, to
compare the grand-canonical results to the behavior of the model in
the canonical situation, we consider that when the polymers are placed
in an excess of solvent (dilute situation), we actually have them
coexisting with the pure solvent phase, which corresponds to the
non-polymerized phase, stable in the grand-canonical solution
for low values of the monomer activity. The density of the coexisting
polymer phase may vanish, and therefore we have a critical situation
associated to extended polymers, or may be finite, corresponding to
collapsed polymers. In the
present calculations, we considered the monomers located on the same
site to be indistinguishable, as was done in the previous work
\cite{oss08}, and in opposition to the calculations done in
\cite{ss07}, where the distinguishable case was treated. One of the
reasons to do so is that the simulations presented in \cite{kpor06}
were also performed for the indistinguishable case. Another
additional point which we address is the location of coexistence
surfaces in the phase diagram of models solved on hierarchical
lattices such as the Bethe lattice. In the previous calculations of
the MMS model on 
treelike lattices, the recursion relations were used directly to find
the coexistence {\em loci} iterating them with
initial conditions defined by the parameters of the model, a procedure
proposed in \cite{p02}. Although this procedure has considerable
physical appeal, it is not granted that the results furnished by it
will be consistent with the ones provided by free-energy
calculations. Actually, in more complex models the `natural' initial
conditions (NIC) may not be unique and there is the possibility that
different locations for the coexistence surface are obtained for
different reasonable choices of these conditions. Therefore, in the
present 
calculations we decided to adopt a definition for the bulk free energy
per site which was proposed some time ago by Gujrati \cite{g95} and
has lead to the same results as other methods based on more solid
foundations, such 
as the integration over the order parameter, in models where it is
possible to perform these calculations. We found that in the present
model the revision of the procedure to find the coexistence surfaces
actually lead to qualitative changes in the phase diagram, thus
showing that the NIC method adopted before may actually lead
to results which are not close to the ones provided by more reliable
calculations. 

In section \ref{defmod} the model is defined in more detail and its
solution on the Bethe lattice is obtained. The
thermodynamic behavior of the model is presented in section
\ref{tpm}. In section \ref{canonico}, we discuss the relation between
the grand-canonical and the canonical behavior of the model, comparing
the results of the present calculation with the finding of the
simulations performed by Krawczyk et al \cite{kpor06}. Final
discussions may be found in section \ref{dc}. Some more results and
discussions concerning the location of coexistence surfaces on Bethe
lattice solutions have been placed in the appendix \ref{ap}, and the
determination of the multicritical point in the parameter space may be
found in the appendix \ref{lmp}.

\section{Definition of the model and solution in terms of recursion relations}
\label{defmod}

\begin{figure}[h!]
\centering
\includegraphics[width=7cm]{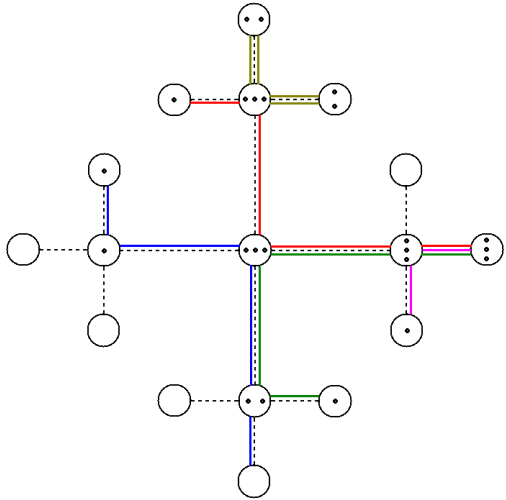}
\caption{(Color online) A contribution to the partition function of
  the model on a Cayley tree with $q = 4$ and 3 generations. The
  weight of this contribution will be $\omega_{1}^{6} \omega_{2}^{3}
  \omega_{3}^{4}$.} 
\label{rede}
\end{figure}

We consider self- and mutually avoiding walks on a Cayley tree with
arbitrary coordination number $q$, imposing the constraint which
forbids immediate reversals. The endpoints of the walks are placed on
the surface of the tree. The grand-canonical partition function of the
model will be given by: 
\begin{equation}
 Y = \sum \omega_{1}^{N_{1}} \omega_{2}^{N_{2}} \omega_{3}^{N_{3}}
\end{equation}
where the sum is over the configurations of the walks on the tree,
while $N_{i}$ , $i = 1, 2, 3$ is the number of sites visited $i$ times
by the walks or, in other words, the number of sites with $i$ monomers
in the configuration. Thus, $\omega_i$, $i=1,2,3$ are the statistical
weights of a site visited $i$ times or, in other words, with $i$
monomers placed on them. In Fig. \ref{rede} a contribution to the
partition function is shown. 

\begin{figure}[h!]
\centering
\includegraphics[width=8cm]{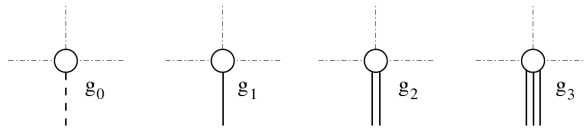}
\caption{Illustration of the rooted subtrees which correspond to the
  partial partition functions.} 
\label{roots}
\end{figure}

As usual, to solve the model on the Bethe lattice we start considering
rooted subtrees of the Cayley tree, defining partial partition
functions for them, where we sum over all possible configurations for a
fixed configuration of the root of the subtree. We thus define four
partial partition functions $g_{i}$ , $i = 0, 1, 2, 3$, where $i$
corresponds to the number of polymer bonds placed on the root edge of
the subtree. The subtrees are shown in Fig. \ref{roots}. 

Actually, it is important to discuss a point about these choice for
the partial partition functions. For the ones with multiple bonds at
the root edge ($g_2$ and $g_3$) a distinction should be made between
the situations where two or more bonds are {\em distinguishable} or
not, since this distinction is important when we will define the
combinatorial coefficients in the recursion relations
below. A pair of bonds which are {\em indistinguishable} corresponds
to the situation where both chains visit the same sequence of sites
since the surface of the tree. Therefore, $g_2$ should be split into
two different partial 
partition functions and three cases should be considered for
$g_3$. Thus, the total number of partial partition functions would be
equal to seven. However, we found out that it is sufficient to
consider only the case where all bonds are distinguishable, since the
ratios of $g_i/g_0$ where $g_i$ is a partial partition function of a
configuration with indistinguishable bonds all vanish at the relevant
fixed 
points of the recursion relations defined below, which corresponds to
the thermodynamic behavior of the model. Actually, there are two fixed
points where the density of edges with one polymer bond on them
vanishes, and for the phases associated to these fixed points (the {\bf
  DO} and {\bf TO} fixed points described below) the inclusion of
indistinguishable polymer bonds in the recursion relations leads to
different results, since
their density does not vanish in the thermodynamic limit. However, we
found that these phases will never be the one with lowest free
energy in any point of the parameter space, and thus if we do not
include indistinguishable bonds in our discussion this will not imply
any change in the phase diagram. Therefore, we decided, for
simplicity, to restrict the discussion of the model to the case of
distinguishable bonds only. In other words, in the limit
of an infinite tree, chains with indistinguishable bonds will never
reach the central region of the tree if the density of edges with
single polymer bonds does not vanish in this region.  This may be
understood
if we notice that in the recursion relations for partial
partition functions with indistinguishable bonds there are always
configurations connecting them to other configurations without those
bonds, provided single bonds are incident at the same site, but never
the opposite happens, so it is not surprising that 
these contributions vanish at the fixed points. We will discuss this
point in some more detail below, when the recursion relations for
the partial partition functions will be obtained. Also, we notice that
if we would adopt that the surface sites of the tree should not be
occupied by two monomers, 
we never would have indistinguishable bonds at any stage of the
iteration, but in treelike structures like the Bethe lattice it is not
granted that such a constraint will not influence the phase diagram of
the model. 

We then proceed obtaining recursion relations for the partial
partition functions, by considering the operation of attaching $q-1$
subtrees with a certain number of generations to a new root site and
edge, thus building a subtree with an additional generation. Below the
recursion relations are presented. In general, we have $g'_{i} =
\sum_{j} g'_{i,j}$ , where the primes denote the partial partition
function on the subtree with one more generation. Whenever
appropriate, the contributions to the sums begin with a product of two
numerical factors, the first of which is the multiplicity of the
configuration of the incoming bonds and the second is the multiplicity
of the connections with the monomers located at the new site. As
discussed in the introduction, we consider the monomers to be
indistinguishable, in opposition to what we did in the particular case
$K = 2$ of the RF model we have studied before
\cite{ss07}. Actually, in the recursion relations below, to
obtain the results for the case of distinguishable monomers one simply
has to replace $\omega_{2}$ by $2 \omega_{2}$ and $\omega_{3}$ by $6
\omega_{3}$. The recursion relation for $g_{0}$ is the sum of the
contributions: 

\begin{subequations}
\begin{eqnarray}
g'_{0,1} &=&   g_0^{q-1}, \\
\mbox{}  \nonumber  \\
g'_{0,2}\,&=&\, \binom{q-1}{2} \times 1 \,  \omega_1 \,g_0^{q-3} \,g_1^2,
\\
\mbox{}  \nonumber  \\
g'_{0,3}\,&=&\, \binom{q-1}{4} \times 3 \, \omega_2 \,g_0^{q-5} \,g_1^4 ,\\
\mbox{}  \nonumber  \\
g'_{0,4}\,&=&\, 3\,\binom{q-1}{3} \times 2   \, \omega_2 \,g_0^{q-4}
\,g_1^2 \,g_2,\\ 
\mbox{}  \nonumber  \\
g'_{0,5}\,&=&\, \binom{q-1}{2} \times 2 \, \omega_2 \,g_0^{q-3} \,g_2^2, \\
\mbox{}  \nonumber  \\
g'_{0,6}\,&=&\,  \binom{q-1}{6} \times 15 \, \omega_3 \,g_0^{q-7} \,g_1^6, \\
\mbox{}  \nonumber  \\
g'_{0,7}\,&=&\, 5 \,\binom{q-1}{5} \times 12 \, \omega_3 \,g_0^{q-6} \,g_1^4 \, 
g_2, \\  
\mbox{}  \nonumber  \\
g'_{0,8}\,&=&\, 6\, \binom{q-1}{4} \times 10 \,\omega_3 \,g_0^{q-5} \,g_1^2 \, 
g_2^2, \\
\mbox{}  \nonumber  \\
g'_{0,9}\,&=&\, \binom{q-1}{3} \times 8 \,\omega_3 \,g_0^{q-4} \,g_2^3, \\
\mbox{}  \nonumber  \\
g'_{0,10}\,&=&\, 4\, \binom{q-1}{4} \times 6 \,\omega_3 \,g_0^{q-5} \,g_1^3\, 
g_3, \\
\mbox{}  \nonumber  \\
g'_{0,11}\,&=&\,  6\, \binom{q-1}{3} \times 6 \,\omega_3 \,g_0^{q-4} \,g_1 \,
g_2 \, g_3, \\
\mbox{}  \nonumber  \\
g'_{0,12}\,&=&\,  \binom{q-1}{2} \times 6 \,\omega_3 \,g_0^{q-3} \,g_3^2, \\
\end{eqnarray}
\end{subequations}
The monomials in the recursion relation for $g'_1$ are:
\begin{subequations}
\begin{eqnarray}
g'_{1,1}\,&=&\, (q-1) \times 1  \, \omega_1 \,g_0^{q-2} \,g_1, \\
\mbox{}  \nonumber  \\
g'_{1,2}\,&=& \binom{q-1}{3} \times 3 \, \omega_2 \, g_0^{q-4}
\,g_1^3 , \\
\mbox{}  \nonumber  \\
g'_{1,3}\,&=&\, 2 \binom{q-1}{2} \times 2 \, \omega_2 \,g_0^{q-3}
\,g_1 \,g_2, \label{1,3} \\
\mbox{}  \nonumber  \\
g'_{1,4}\,&=&\,  \binom{q-1}{5} \times 15 \, \omega_3 \,g_0^{q-6} \,g_1^5, \\
\mbox{}  \nonumber  \\
g'_{1,5}\,&=&\, 4 \binom{q-1}{4} \times 12 \,\omega_3 \,g_0^{q-5} \,g_1^3 \, 
g_2, \\
\mbox{}  \nonumber  \\
g'_{1,6}\,&=&\, 3 \binom{q-1}{3} \times 10\, \omega_3 \,g_0^{q-4} \,g_1 \, 
g_2^2, \\
\mbox{}  \nonumber  \\
g'_{1,7}\,&=&\,  3 \binom{q-1}{3} \times 6\,\omega_3 \,g_0^{q-4} \,g_1^2 \, 
g_3, \\
\mbox{}  \nonumber  \\
g'_{1,8}\,&=&\,  2 \binom{q-1}{2} \times 6\,\omega_3 \,g_0^{q-3} \,g_2\,g_3\,.
\end{eqnarray}
\end{subequations}
Let us  illustrate the differences in the recursion relations when
indistinguishable bonds are included considering, for instance, the
contribution \ref{1,3} above. In this contribution, a edge with two
bonds  and another with a single bond reach the root site from above,
and a single polymer bond proceeds to the root edge. Therefore, one of
the polymer bonds of the double bonded edges is connected to the bond
at the root, while the other one is linked to the other bond coming
from above. If the bonds in the double bonded incoming edge are
distinguishable, there are two distinct linking configurations, thus
leading to the second factor 2 in the recursion relation. If
the polymer bonds on the same edge were indistinguishable, this factor
would be unitary, and it is worth noting that this contribution would
end a chain of double bonds which has started at the surface of the
tree, splitting it into two edges with single bonds.

For $g'_2$ we find the contributions:
\begin{subequations}
\begin{eqnarray}
g'_{2,1}\, &=& \binom{q-1}{2} \times 1 \, \omega_2 \,g_0^{q-3} \,g_1^2,\\
\mbox{}  \nonumber  \\
g'_{2,2}\,&=&\, (q-1) \times 1 \, \omega_2  \,g_0^{q-2} \,g_2 , \\
\mbox{}  \nonumber  \\
g'_{2,3}\,&=&\,\binom{q-1}{4} \times 6 \, \omega_3 \,g_0^{q-5} \,g_1^4, \\
\mbox{}  \nonumber  \\
g'_{2,4}\,&=&\, 3 \binom{q-1}{3} \times 5 \, \omega_3 \,g_0^{q-4} \,g_1^2 \, 
g_2, \\
\mbox{}  \nonumber  \\
g'_{2,5}\,&=&\, \binom{q-1}{2} \times 4 \,\omega_3 \,g_0^{q-3} \, g_2^2, \\
\mbox{}  \nonumber  \\
g'_{2,6}\,&=&\,  2 \binom{q-1}{2} \times 3 \, \omega_3 \,g_0^{q-3} \,g_1\,g_3\,.
\end{eqnarray}
\label{rrg2rf}
\end{subequations}

Finally, the contributions to the recursion relations for $g_3'$ are

\begin{subequations}
\begin{eqnarray}
g'_{3,1}\, &=& \binom{q-1}{3} \times 1 \, \omega_3 \,g_0^{q-4} \,g_1^3,\\
\mbox{}  \nonumber  \\
g'_{3,2}\,&=&\,  2 \binom{q-1}{2} \times 1 \,\omega_3  \,g_0^{q-3} \,g_1 \, 
g_2 , \\
\mbox{} \nonumber  \\
g'_{3,3}\,&=&\, (q-1)\times 1 \,  \omega_3 \,g_0^{q-2} \,g_3.
\end{eqnarray}
\label{rrg3rf}
\end{subequations}

The partial partition functions often grow exponentially with the number
of iterations , so that
we may now define ratios of the partial partition functions $R_1=g_1/g_0$,
$R_2=g_2/g_0$. and  $R_3=g_3/g_0$, and write the recursion relations 
for these ratios, which usually remain finite in the thermodynamic limit.
The results are:
\begin{subequations}
\begin{eqnarray}
R'_1 \,&=&\, \frac{1}{D} \, \left[  (q-1) \, \omega_1 \, R_1 \,+\,
 \, 3 \binom{q-1}{3} \,\omega_2 \, R_1^3 \,+\,
\right. \nonumber \\ 
&&  4 \binom{q-1}{2}\,  \omega_2 \,R_1\,R_2\,+\,
15 \binom{q-1}{5} \,\omega_3 \, R_1^5 \,+\,
48 \binom{q-1}{4} \,\omega_3 \, R_1^3 \,R_2 \,+   \nonumber \\
&& \left. 30\,\binom{q-1}{3}  \,\omega_3 \, R_1 \,R_2^2 \,+\,
18 \,\binom{q-1}{3}  \,\omega_3 \, R_1^2 \,R_3 \,+\,
12 \, \binom{q-1}{2}  \,\omega_3 \, R_2 \,R_3 \right] \,,   
\label{rrr1}\\
R'_2 \,&=& \,\frac{1}{D} \,\left[\binom{q-1}{2} \,\omega_2\,R_1^2 \,+\,
 (q-1) \,\omega_2\, R_2 \,+\,
6 \,  \binom{q-1}{4} \,\omega_3 \, R_1^4 \,+ \right.  \nonumber \\
&& \left. 15 \,\binom{q-1}{3}  \,\omega_3 \, R_1^2 \,R_2 \,+\,
4 \, \binom{q-1}{2}\, \omega_3 \, R_2^2 \,+\,
6 \,\binom{q-1}{2}  \,\omega_3 \, R_1 \,R_3 \right]\,,   \label{rrr2} \\ 
R'_3 \,&=& \,\frac{\omega_3}{D} \,\left[ \binom{q-1}{3} \, R_1^3 \,+\,
2  \, \binom{q-1}{2}  \, R_1 \,R_2 \,+\,(q-1) \, R_3
\right]\,. \label{rrr3} 
\end{eqnarray}
\label{rrrf}
\end{subequations}
where
\begin{eqnarray}
D\,&=&\,1\,+\, \binom{q-1}{2} \omega_1  \,R_1^2 \,+\,
 \, 3 \binom{q-1}{4}\,\omega_2 \,R_1^4 \, +\,
6 \binom{q-1}{3}\,\omega_2  \,R_1^2 \,R_2\,+ \nonumber \\
&& 2 \binom{q-1}{2}\,\omega_2  \,R_2^2 \,+\,
15 \, \binom{q-1}{6}\, \omega_3 \, R_1^6  \, +\,
60 \, \binom{q-1}{5}\, \omega_3 \, R_1^4 \,R_2  \, + \nonumber \\
&& 60 \, \binom{q-1}{4}\, \omega_3 \, R_1^2 \,R_2^2  \, + \,
8  \, \binom{q-1}{3}\, \omega_3 \, R_2^3  \, + \,
24 \, \binom{q-1}{4}\, \omega_3 \, R_1^3 \,R_3  \, + \nonumber \\
&&36  \, \binom{q-1}{3}\, \omega_3 \, R_1 \,R_2 \,R_3  \, + \,
6 \, \binom{q-1}{2}\, \omega_3 \, R_3^2 \,.
\label{denom}
\end{eqnarray}

The partition function of the model on the Cayley tree may be obtained
if we consider the operation of attaching $q$ subtrees to the central
site of the lattice. The result is:

\begin{eqnarray}
Y \,&=&\, g_0^q+\binom{q}{2}\,\omega_1\,g_0^{q-2}\,g_1^2\,+\,
3\binom{q}{4}\,\omega_2\,g_0^{q-4}\,g_1^4\,+ \,\
6 \binom{q}{3}\,\omega_2\,g_0^{q-3}\,g_1^2\,g_2\,+ \nonumber \\
&&  2 \binom{q}{2}\,\omega_2\,g_0^{q-2}\,g_2^2\,+ \,
15 \,\binom{q}{6}\, \omega_3 \,  \,g_0^{q-6} \,g_1^6, \,+\,
60  \,\binom{q}{5} \, \omega_3 \,g_0^{q-5} \,g_1^4 \, g_2  \,+\nonumber \\
&& 60 \, \binom{q}{4} \,\omega_3 \,g_0^{q-4} \,g_1^2 \, g_2^2\,+ \,
8 \,  \binom{q}{3} \,\omega_3 \,g_0^{q-3} \,g_2^3 \,+ \,
24 \, \binom{q}{4}  \,\omega_3 \,g_0^{q-4} \,g_1^3\, g_3 \,+ \nonumber \\
&& 36 \, \binom{q}{3}  \,\omega_3 \,g_0^{q-3} \,g_1 \,g_2 \, g_3  \,+ \,
6 \, \binom{q}{2} \,\omega_3 \,g_0^{q-2} \,g_3^2 \,.
\end{eqnarray}

Using the partition function above, we then proceed
calculating the densities at the central site of the tree. The density
of monomers is given by:
\begin{equation}
\rho=\rho_1+\rho_2+\rho_3=\frac{P}{T}+\frac{2Q}{T}+\frac{3S}{T},
\label{dm}
\end{equation}
where:
\begin{equation}
T=1+P+Q+S,
\end{equation}
and
\begin{subequations}
\begin{eqnarray}
P \,&=&\,\omega_1 \, \binom{q}{2}\,R_1^2, \\
Q \,&=&\,\omega_2\,\left[ 3\binom{q}{4}\,R_1^4\,+
\,6 \binom{q}{3}\,R_1^2\,R_2\,+\,
2 \binom{q}{2}\,R_2^2 \right]. \\
S \,&=&\,\omega_3 \,\left[ 
15 \,\binom{q}{6}\, R_1^6, \,+\,
60  \,\binom{q}{5} \, R_1^4 \, R_2  \,+ \right.\nonumber \\
&& 60 \, \binom{q}{4}  \,R_1^2 \, R_2^2\,+ \,
8 \,  \binom{q}{3}  \,R_2^3 \,+ \,
24 \, \binom{q}{4}  \,R_1^3\, R_3 \,+ \nonumber \\
&& \left. 36 \, \binom{q}{3}   \,R_1 \,R_2 \, R_3  \,+ \,
6 \, \binom{q}{2}  \,R_3^2 
\right].
\end{eqnarray}
\label{pqrf}
\end{subequations}

\section{Thermodynamic properties of the model}
\label{tpm}

\subsection{Fixed points}

The thermodynamic phases of the system on the Bethe lattice will be
given by the stable fixed points of the recursion relations, which are
reached after infinite iterations and thus correspond to the
thermodynamic limit. We find five different fixed points, which are
described below: 

\begin{enumerate}
\item{Non-polymerized (\textbf{NP}) fixed point:}

This fixed point is characterized by $R_{1}^{NP} = R_{2}^{NP} =
R_{3}^{NP} = 0$, and therefore all densities vanish. In order to study
the stability region in the parameter space for this fixed point, we
consider the Jacobian: 
\begin{equation}
J_{i,j}=\left.\frac{\partial R'_i}{\partial R_j}\right|_{R_1=R_2=R_3=0}\,=\,
\left( \begin{array}{cccc}
(q-1) \,\omega_1 &  0  & 0  \\ \\
0  & (q-1)\,\omega_2  & 0\\ \\
0 & 0 & (q-1) \, \omega_3
\end{array}
\right) \;,
\end{equation}
and the region of the parameter space for which the largest eigenvalue
of the Jacobian is smaller than one and therefore the \textbf{NP}
phase is stable, is the one for which the three inequalities below are
simultaneously satisfied: 
\begin{equation}
\label{enp}
\omega_1 \,<\, \frac{1}{q-1} \;\;\;;\;\;\; \omega_2 \,< \, \frac{1}{q-1}
\;\;\;;\mbox{and} \;\;\;  \omega_3 \,< \, \frac{1}{q-1}.
\end{equation}

\item{Double occupancy (\textbf{DO}) fixed point:}

In this fixed point the ratios are given by $R_{1} = R_{3} = 0$ and
$R_{2} = R_{2}^{DO} \neq 0$. The fixed point value of $R_{2}$ will be
one of the solutions of the cubic equation: 
\begin{equation}
\label{cdo}
8  \, \binom{q-1}{3}\, \omega_3 \, [R_2^{DO}]^3\,+\, 2 \, \binom{q-1}{2}\,
\omega_2 [R_2^{DO}]^2\,-\,4 \, \binom{q-1}{2}\,\omega_3 \,R_2^{DO} 
\,-\, (q-1)\,\omega_2\,+\,1
\,=\,0.
\end{equation}
For $q = 3$ the cubic term of the equation vanishes and a simple
expression is found for the fixed point value of $R_{2}$: 
\begin{equation}
\label{r2doq3}
R_{2;q=3}^{(DO)}\,=\,\frac{2\, \omega_3 \,\pm\,\sqrt{(4 \, \omega_3)^2
    + 2\, ( 2
\, \omega_2 -1)  \omega_2}}{2 \, \omega_2}.
\end{equation}

It is worth noticing that this fixed point does not disappear when
$\omega_{2}=0$, $\omega_{3} \neq 0$. Actually, in general it
corresponds to a double occupancy of \textit{bonds} and not
necessarily of sites. Also, it is easy to obtain the elements of the
Jacobian at this fixed point, as a function of the statistical weights
and $R_2^{DO}$.

\item{Triple occupancy (\textbf{TO}) fixed point:}

At this fixed point, we have $R_{1} = R_{2} = 0$ ; $R_{3} = R_{3}^{TO}
\neq 0$, and the fixed point value of the ratio $R_{3}$ is given by: 
\begin{equation}
\label{pfto}
R_3^{TO}\,=\, \sqrt{\frac{(q-1)\,\omega_3\,-\,1}{3 (q-1) (q-2)
    \,\omega_3}}, 
\end{equation}
The Jacobian for this fixed point will be
\begin{equation}
J^{(TO)}_{i,j}=\left.\frac{\partial R'_i}{\partial R_j}\right|_{TO}\,=\,
\left( \begin{array}{cccc}
\frac{\omega_1}{\omega_3} &  \sqrt{\frac{3\, (q-2)\,[(q-1)\,\omega_3-1]}
{(q-1)\,\omega_3}}  & 0  \\ \\
\, \sqrt{\frac{12 (q-2)\,[(q-1)\,\omega_3-1]}{(q-1)\,\omega_3}}   &
\frac{\omega_2}{\omega_3}  & 0\\ \\
0 & 0 & \frac{2}{(q-1) \, \omega_3}-1
\end{array}
\right) \;.
\end{equation}

We found two additional fixed points which display all ratios
different from zero. There exists a region in the parameter space
where both are stable, thus a coexistence surface of both regular
polymerized phases is found, as will be seen below. The two phases
are: 

\item{Regular polymerized (\textbf{P1}) fixed point:}

This phase is stable in a region situated at $\omega_1>\frac{1}{q-1}$
and for small values of $\omega_3$

\item{Regular polymerized (\textbf{P2}) fixed point:}

This fixed point is stable for sufficiently large values of
$\omega_3$. At the coexistence surface of both regular polymerized
phases, \textbf{P1} is more anisotropic than \textbf{P2}, in the sense
that in the former $R_1,R_2 \gg R_3$ and $\rho_1 \gg \rho_2,\rho_3$,
while in phase \textbf{P2} ratios and densities are more balanced. 

\end{enumerate}

In some fixed points (mainly in the last two), we were unable to
perform an analytic study of
the Jacobian as a function of the statistical weights, but it is
easy to obtain numerically the matrix elements as functions of these
weights and 
the fixed point values of the ratios. In this way, we obtained the
stability limits (spinodals) of the 
five fixed points (or phases), in order, to characterize the
transitions between them. As we are dealing with three
parameters, the spinodals are surfaces in the parameter space
($\omega_{1},\omega_{2},\omega_{3}$). The continuous transitions
(surfaces, lines or points) happen in the regions where the spinodals
of the different phases are coincident. The coexistence surfaces are
bounded by the spinodals, but for their precise location in the
parameter space it is necessary to obtain the bulk free energy of the
Bethe lattice solution. 

\subsection{Free energy}

It is useful, particularly to find the coexistence surfaces in the
phase diagrams, to calculate the free energies of the various
thermodynamic phases of the model. One possibility would be to perform
the Maxwell construction, and actually this was done for similar
models some time ago \cite{sms96}, but this procedure would be awkward
in the present case, particularly in regions of the parameter space
where more than two fixed points are stable. A simple
way to find the coexistence region \cite{p02} is just to iterate the
recursion relations starting with `natural' initial conditions, that is,
using initial values for the ratios which correspond to a reasonable
choice for the configurations at the surface of the Cayley
tree. Actually, we used this procedure in our recent works on the $K =
2$ case of the present model \cite{ss07,oss08}. Although this procedure
has a considerable physical appeal, is simple and leads to reasonable
results, we were not able to justify it starting from basic
principles. In particular, for more complex models, the `natural'
initial conditions may not be unique, and different choices for them
could lead to different results for the locus of coexistence. On the
other side, the Maxwell construction follows directly from the
recursion relations, being therefore independent from the choice of
initial conditions for the iterations. So we decided to use a
procedure described by Gujrati \cite{g95} to find the free energy of
the thermodynamic phases. 

We will briefly discuss Gujrati's argument, in a version appropriate
for the present model. We consider the grand-canonical free energy of
the model on the Cayley tree with $M$ generations $\tilde{\Phi}_{M} =
−k_{B} T \ln Y_{M}$ , where $T$ is the 
absolute temperature and $k_{B}$ stands for Boltzmann's constant. We
may then define a reduced adimensional free energy $\Phi_{M} =
\tilde{\Phi}_{M} / k_{B} T$. It is usual in finite size scaling
arguments to consider the free energy per site on regular lattices to
be different for sites on the surface and in the bulk of the
lattice. Here we will make a similar ansatz, and suppose the reduced
free energy per site to be the same for all sites of the same
generation of the Cayley tree. Let us number the generations starting
at $0$ for the surface of the tree, and define $\phi_{0}$ as the
reduced free energy per site for the $q (q - 1)^{M-1}$ sites on the
surface of the $M$-generations tree, $\phi_{1}$ will be the free
energy per site for the $q (q - 1)^{M-2}$ sites the first generation
and so on. We may then write the total free energy of the Cayley tree
as: 
\begin{equation}
\Phi_{M} = q(q - 1)^{M - 1} \phi_{0} + q(q - 1)^{M - 2} \phi_{1} +
. . . + q \phi_{M-1} + \phi_{M} , 
\end{equation} 
where $\phi_{M}$ is the reduced free energy of the central site of the
tree, which is the one which should correspond to 
the Bethe lattice solution of the model. We may now write a similar
expression for a tree with $M + 1$ generations, 
assuming a homogeneity condition which states the free energy per site
for sites of the same generation of the two 
trees to be the same, so that:

\begin{equation}
\Phi_{M+1} = q(q - 1)^{M} \phi_{0} + q(q - 1)^{M - 1} \phi_{1} +
. . . + q \phi_{M} + \phi_{M+1} , 
\end{equation}
By inspection, we may readily realize that $\Phi_{M +1} - (q - 1)
\Phi_{M} = \phi_{M} + \phi_{M+1}$. If we now consider that in the 
thermodynamic limit $M \rightarrow \infty$ the free energies of the
central sites of both trees should be the same, that is $\phi_{M+1} =
\phi_{M} = \phi_{b}$, where $\phi_{b}$ is the bulk free energy per
site which corresponds to the Bethe lattice solution, we have
$\Phi_{M+1} - (q - 1) \Phi_{M} = 2\phi_{b}$, so that the reduced free
energy per site for the Bethe lattice is: 
\begin{equation}
 \phi_{b} = - \frac{1}{2} \lim_{M \rightarrow \infty} \ln \left(
 \frac{Y_{M+1}}{Y_{M}^{(q-1)}} \right).
\end{equation}
This result is equivalent to the expression (3) in reference
\cite{g95}, although here we are considering a less general situation
than the original work. 

Evaluating the ratio of partition functions in the expression above
for the reduced bulk free energy at the fixed point, we find that: 
\begin{equation}
 \lim_{M \rightarrow \infty} \frac{Y_{M+1}}{Y_{M}^{(q-1)}} =
 \frac{D^{q}}{(1+P+Q+S)^{q-2}}  
\end{equation}
and therefore we have:
\begin{equation}
 \phi_{b} = - \frac{1}{2} \left[ q \ln D - (q-2) \ln y \right]  
\label{bfe}
\end{equation}
where $y = 1 + P + Q + S$ and $D$ are calculated at the fixed point $M
\rightarrow \infty$.

\subsection{Phase diagrams}

Using the spinodals to find the continuous transitions and the free
energy to determine the coexistence surfaces we are able to build the
whole phase diagram of the system. We will show some cuts of the phase
diagram for $q=4$, as well as a perspective of the whole
diagram in the three-dimensional parameter space.

\begin{figure}[h!]
\centering
\includegraphics[width=8cm]{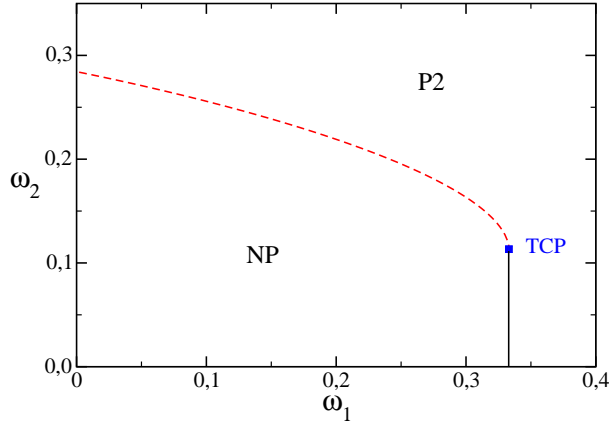}
\caption{(Color online) Phase diagram for $\omega_{3}=0$. The dashed
  (red) curve is a first order transition and the full (black)
  line is a continuous transition between the NP and P
  phases. The tricritical point TCP, represented by a square (blue),
  separates the two transition
  lines. This and the following diagrams were all obtained for $q=4$.}
\label{w1w2}
\end{figure}

The diagram for $\omega_{3}=0$ ($K=2$) is shown in
Fig. \ref{w1w2}. For 
small values of $\omega_{2}$ we find a continuous transition, between
the phases NP and P2, which ends at a tricritical point (TCP) located
at $\omega_{1}^{TCP}=\frac{1}{(q-1)}$ and
$\omega_{2}^{TCP}=\frac{1}{(q-1)^{2}}$, as found in \cite{ss07}. Above
the tricritical point the transition becomes discontinuous. Here it is 
important to stress that this particular case ($\omega_{3}=0$) was
studied in \cite{ss07}, considering
distinguishable monomers and using the NIC
method to find the coexistence lines. There three
``stable`` phases were found: NP, P and DO, however, here (using the
free 
energies) we find only two stable phases: NP and P2. Indeed, the DO
fixed point is stable in a region of the parameter space, but the
corresponding phase is never the one with the lowest free energy, that
is, its free energy is always greater than that of the phases P2
or NP. In the same way, the TO phase is never the most stable in
any region of the phase diagram, although, if we use the NIC method it
appears to be stable for small values of $\omega_{1}$ and $\omega_{2}$
and large $\omega_{3}$. A detailed discussion of this point may be
found the appendix \ref{ap}. 

\begin{figure}[h!]
\centering
\includegraphics[width=8cm]{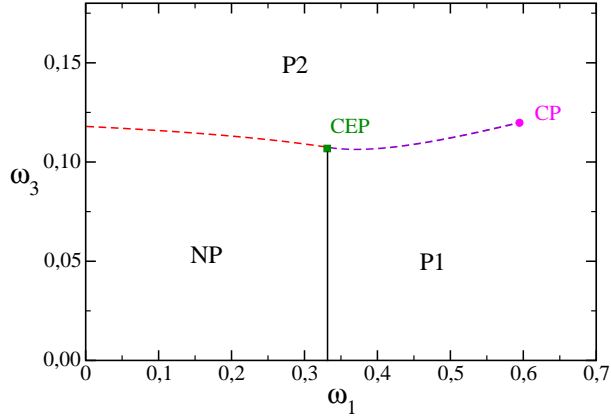}
\caption{(Color online) Phase diagram for $\omega_{2}=0$. The dashed
  curve located left of the critical endpoint (CEP) (red) is a
  coexistence line between the phases NP and P2, and at the dashed
  curve right of the CEP (violet) phases P1 and P2 coexist. These two
  phases become indistinguishable at the critical point (CP)
  represented by a (purple) circle. Phases NP and P1 are separated by
  a continuous transition, represented by the (black) full line. This
  line meets the coexistence line at a critical endpoint (CEP),
  represented by a (green) square.}
\label{w1w3}
\end{figure}

\begin{figure}[h!]
\centering
\includegraphics[width=8cm]{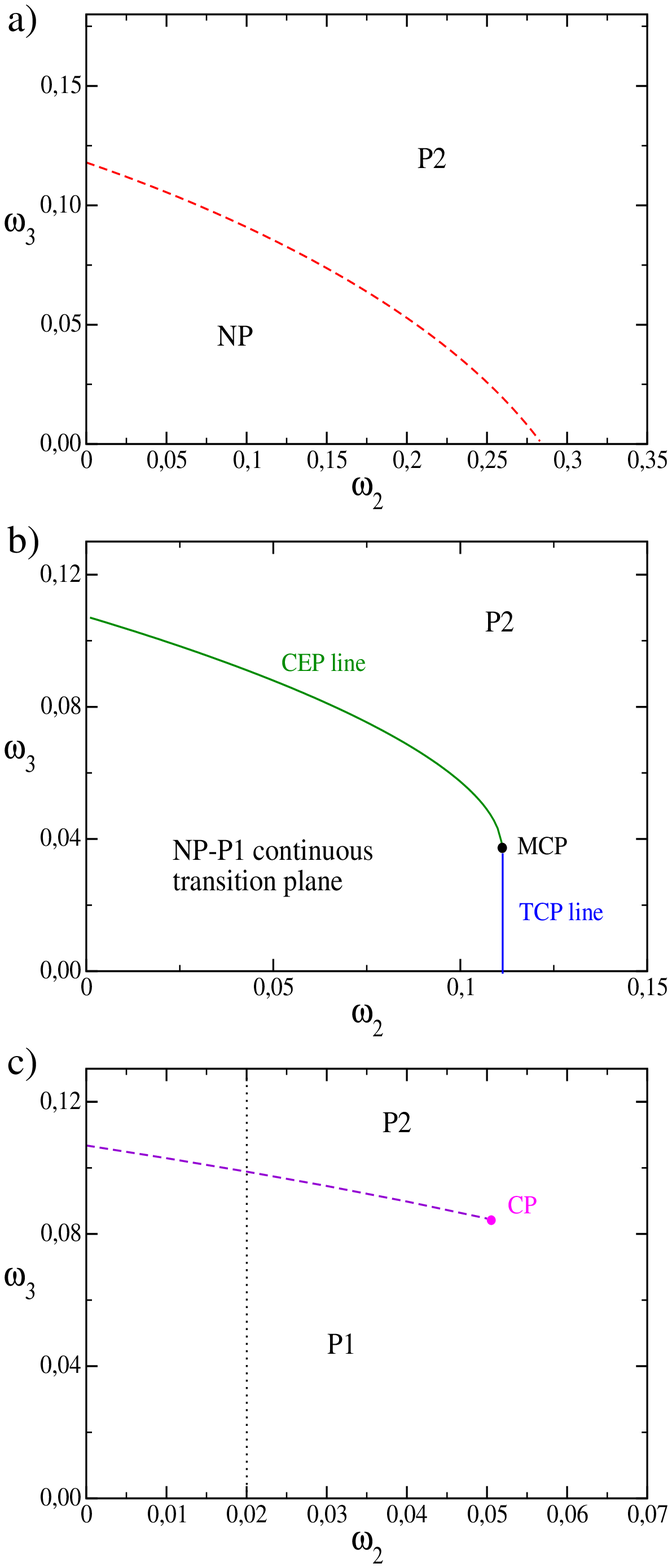}
\caption{(Color online) Phase diagrams for: a) $\omega_{1}=0$, b)
  $\omega_{1}=1/3$ and c) $\omega_{1}=0.40$. The dashed curves are  
  discontinuous transitions between phases NP and
  P2 (red) in a) and phases P1 and P2 (violet) in c). At the dotted line
  in c) the densities shown in Fig. \ref{densi} were calculated.}
\label{w2w3}
\end{figure}

In the $\omega_{2} = 0$ plane, we found a very rich phase diagram, as
is shown in Fig. \ref{w1w3} For $\omega_{1} < \frac{1}{(q-1)}$ we find
a first 
order transition between the phases NP and P2. At
$\omega_{1}=\frac{1}{(q-1)}$, there is a continuous transition line
between the phases NP em P1, this critical line ends at a critical
endpoint (CEP). In the $\omega_{1} > \frac{1}{(q-1)}$ region we have a 
discontinuous transition between the phases P1 and P2 and this
coexistence line ends at a critical point (CP). 

\begin{figure}[h!]
\vspace{1.0cm}
\centering
\includegraphics[width=8cm]{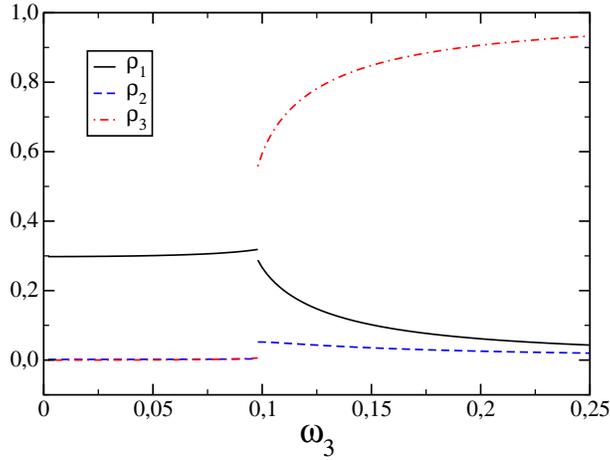}
\caption{(Color online)Densities as functions of $\omega_3$
for $\omega_1=0.4$ and $\omega_2=0.02$. Notice values at the
coexistence of phases P1 and P2.} 
\label{densi}
\end{figure}

In Fig. \ref{w2w3}, we show several diagrams, in the
($\omega_{2},\omega_{3}$) 
space, for different values of $\omega_{1}$. For $\omega_{1}=0$
(a) there is only a single coexistence line between the NP and
P2 phases. Similar diagrams are obtained for all
$\omega_{1} < \frac{1}{(q-1)}$. For $\omega_{1}=\frac{1}{(q-1)}$ (b)
we 
have a critical surface (continuous transition) separating the NP and
P1 phases. This surface is limited by a critical endpoint 
line and a tricritical line, and these two lines meet at a
multicritical point (MCP). The multicritical point is located at
$\omega_{1}^{MCP}=\frac{1}{(q-1)}$,
$\omega_{2}^{MCP}=\frac{1}{(q-1)^{2}}$ and
$\omega_{3}^{MCP}=\frac{1}{(q-1)^3}$, its location is determined in
the appendix \ref{lmp}. The tricritical point line lies
at constant $\omega_{1}=\omega_{1}^{MCP}$ and
$\omega_{2}=\omega_{2}^{MCP}$, and $0 \leq \omega_{3} \leq
\omega_{3}^{MCP}$. For $\omega_{1} > \frac{1}{(q-1)}$ (c) we have a
discontinuous transition between the phases P1 and P2 and this
coexistence 
surface ends at a critical line. The critical
line starts at the multicritical point and the value of $\omega_{2}$
at the line decreases as $\omega_1$ and $\omega_3$ increase, so that
the P1-P2 coexistence surface 
ends at $\omega_{1} = 0.608762(1)$, $\omega_{2} = 0$ and $\omega_{3} =
0.121132(1)$. Also, to illustrate the discussion of the differences in
both regular polymerized phases, we show in Fig. \ref{densi} the
fixed point values of the densities as functions of
$\omega_3$ for $\omega_1=0.4$ and $\omega_2=0.02$ (dashed line in
phase diagram c) in Fig. \ref{w2w3}). 

\begin{figure}[h!]
\centering
\includegraphics[width=12cm]{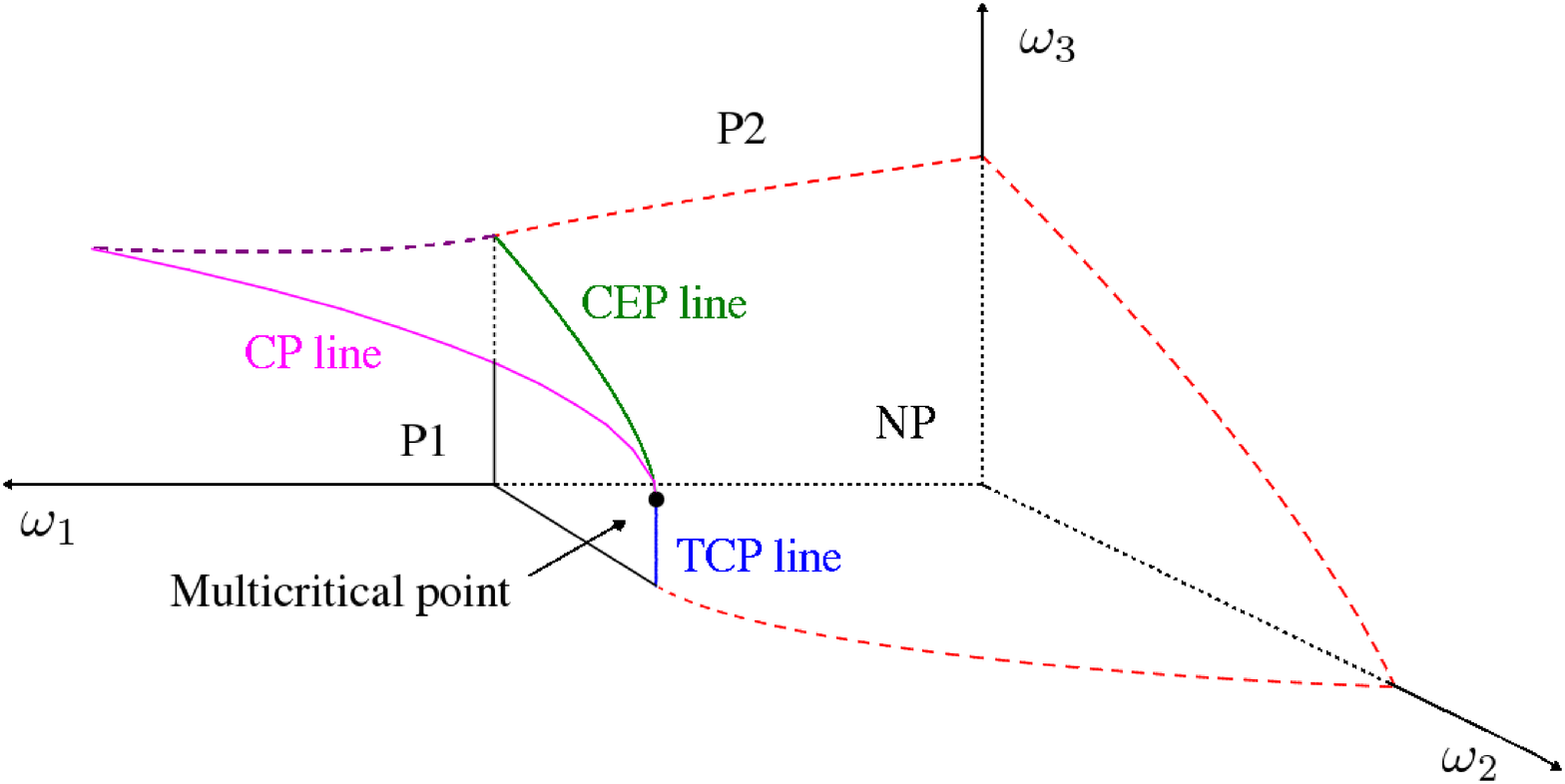}
\caption{(Color online) Sketch of the three-dimensional phase diagram. The
  first order transition surfaces: NP-P2 (red) and P1-P2 (violet),
  limited by dashed lines, are shown.} 
\label{diag3d}
\end{figure}

A sketch of the whole three-dimensional phase diagram is shown in
Fig. \ref{diag3d} 
and this summarizes all the features discussed in the two-dimensional
cuts of the phase diagram presented above. 

\section{Comparison with the canonical simulations}
\label{canonico}

The $K=3$ version of the model were originally studied by Krawczyk et
al. \cite{kpor06} using Monte Carlo simulations. The simulations were
performed considering a chain of constant size placed on an infinite
lattice, so that they are in the canonical ensemble. To compare our
grand-canonical results with the simulations we have to map our
phase diagram to the canonical one. In this particular case, the usual
procedure does not work because in the canonical simulation, as well
as in experiments with diluted chains in a solvent, the canonical
system is not homogeneous, but is composed by isolated chains in an
excess of solvent (empty lattice sites). Therefore, we may say that in the
simulations the polymer coexists with the empty lattice, namely, we
have two phases coexisting: one of them polymerized (the polymer
itself and the empty lattice sites close to it) and a 
non-polymerized (the remaining empty lattice sites). It follows that
in our grand-canonical calculations the canonical situation of the
simulations corresponds to the critical and coexistence surfaces
limiting the NP phase, and the critical lines and points at these
surfaces must be the critical lines in the canonical diagram. 

In the canonical simulations, Krawczyk et al. fixed the energy of a
single 
monomer to be equal to zero, i. e., the Boltzmann weight
$e^{\beta_{0}}=1$. The parameters in the simulations
were $\beta_{\ell}=-\beta \varepsilon_{\ell}$, with $\ell=1,2$, where
$\beta=1/k_BT$ and
$\varepsilon_{\ell}$ is the energy associated with sites occupied by
$\ell+1$ monomers. To relate the simulational parameters to the ones
used in our grand-canonical calculations, we notice that in our
calculations the statistical weight of a site occupied by a single
monomer is $\omega_1=z$, where $z$ is the activity of a monomer. A
energy $\varepsilon_{1}$ is associated to a site with two monomers,
thus the corresponding statistical weight is
$\omega_2=z^2\,e^{-\beta\varepsilon_1}$ and for a site with three
monomers, we have $\omega_3=z^3\,e^{-\beta\varepsilon_2}$. Therefore,
the parameters used in the canonical simulations relate to the
parameters used here as
\begin{subequations} 
\begin{eqnarray}
\beta_1 &=& \ln{\left[ \frac{\omega_{2}}{\omega_{1}^{2}}\right] },\\
\beta_2 &=& \ln{\left[ \frac{\omega_{3}}{\omega_{1}^{3}}\right] }.
\end{eqnarray}
\label{beta}
\end{subequations}

\begin{figure}[h!]
\vspace{1.0cm}
\centering
\includegraphics[width=8cm]{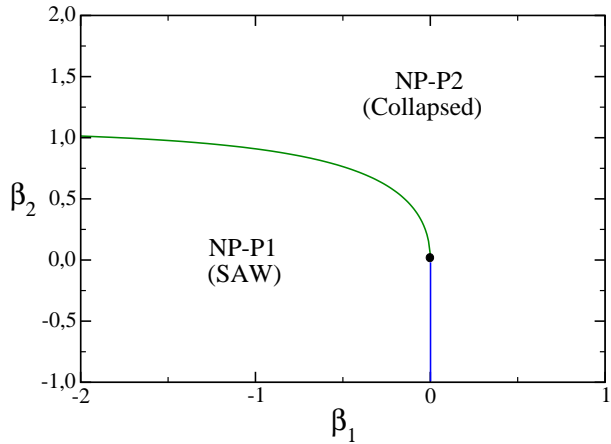}
\caption{(Color online) Canonical phase diagram. The curve at
  $\beta_2>0$ (green) is the
  CEP line and the straight line at $\beta_2<0$ (blue) is the
  tricritical line. The multicritical point, where the two lines meet,
  is located on the origin and represented by a circle.}  
\label{dfcan}
\end{figure}

The canonical phase diagram which we found in the $\beta_{1},
\beta_{2}$ parameters is shown in Fig. \ref{dfcan}, for a Bethe
lattice with $q=4$. The 
multicritical point is located on the origin. The
tricritical line is placed at $\beta_{1}=0$ and $\beta_{2}$ ranging
between $-\infty$ and $0$. The critical endpoint line is in the
negative $\beta_{1}$ and positive $\beta_{2}$ quadrant. For
$\beta_{1} < 0$ and below the CEP curves, we have a region
corresponding to the critical surface NP-P1 in the grand-canonical
phase diagram. Thus, in this region the polymers are formed by chains
with predominance of single visited sites, this 
is agreement with the simulations results. The other
region corresponds to coexistence surface between NP and P2 phases,
thus in this region sites with two or three monomers are more frequent,
characterizing a ``collapsed'' phase. Indeed, a dense polymerized
phase, in the grand-canonical ensemble, is characterized by the
lattice completely occupied by monomers ($\rho = 1$), and, in our
solution method, 
this would appear as a fixed point with one or more diverging ratios
$R_i$. In this model we do not 
find any collapsed phase in this sense, such a phase may be present in
SASAW's with attractive interactions between bonds in elementary squares
on the square lattice \cite{sms96}. Nevertheless, as the density of
the polymerized phases at the coexistence {\em loci} is nonzero, this
already sets $\nu=1/d$ and therefore it is appropriate to call this a
collapsed phase.

The location of the multicritical point in the Bethe lattice
solution ($\beta_{1}=\beta_{2}=0$) is 
not physically reasonable because it corresponds to non-interacting
monomers. We expect that the transition to the collapsed
phase occurs in a attractive region for at least one of the parameters
($\beta_{1},\beta_{2}$), but for $\beta_{2} < 0$ the transition line
is along the $\beta_{2}$ axis ($\beta_{1}=0$) and this it is not
reasonable. But, it is important to keep in mind that our solution is
a mean-field approximation, which generally overestimates the domain of
the ordered phase. Actually, this inconsistency was already noted in
the solution of the case $K=2$ \cite{ss07}, and was one of the
motivation to perform a calculation of this model on the Husimi
lattice, which is expected to lead to results closer to the ones on
regular lattices. This solution, which corresponds to the particular
case  
$\omega_{3}=0$ and $\beta_{2}=-\infty$ of the model considered here,
on the Husimi lattice build 
with squares (a second order approximation for the square lattice)
displays a tricritical point located at $\omega_{1}=0.3325510(6)$ and
$\omega_{2}=0.120544(4)$, which in the canonical variables corresponds
to 
$\beta_{1} \approx 1.09$ and $\beta_{2}=-\infty$. Thus, within this
(better) approximation, at least in $\beta_{2}=-\infty$ limit, we find
the 
transition in the expected region ($\beta_{1} > 0$). This suggests that
the whole tricritical line may be at positive values of $\beta_1$, as
found 
in the simulations, and therefore, the multicritical point may not be
at the origin. 

Finally, there is the question of the order of the transitions. In the
simulations, Krawczyk et al. suggest, by estimating the fluctuations
of the order parameter in their simulations, 
that for $\beta_{1} < 0$ the transition is continuous and for
$\beta_{2} < 0$ it becomes of first-order one, and this lines match at a
tricritical point. In our phase diagram, all transitions are
continuous but we have a tricritical line in $\beta_{2} < 0$ and a CEP
line in $\beta_{1} < 0$ region. Although both transitions are
always continuous the critical exponents should be different in the
two lines, due to the fact that the transitions are of a different
nature (critical endpoints and tricritical points). Actually, since in
the canonical conditions the tricritical line is always approached in
the {\em weak} direction \cite{gw70}, that is, staying on the
coexistence surface, the weak tricritical exponents will be found. In
particular, in three dimensions, the tricritical exponents will be
classical, apart from logarithmic corrections \cite{ls84}. In
particular, as in the simulations  Krawczyk et al. have estimated the
fluctuations of order-parameter like variables, for the tricritical
line the expected exponent would be $\gamma=2$. The critical endpoint
line, however, is characterized by regular critical exponents, and
estimates of $\gamma$ for the polymerization transition in three
dimensions are around $1.158$ \cite{g00}, while the classical value is
equal to unity. The estimates from the
simulations have lead  Krawczyk et al. to suggest that the transition
line which corresponds to the tricritical line in our approximate
calculations should be of first order. They also remark that this
transition appears to be stronger than the one which
corresponds to our critical endpoint line. It remains an open question
if the stronger singularity observed in the former transition could
not be due to the larger exponent for the singularity in the
fluctuations of the order parameter.

\section{Final discussions and conclusion}
\label{dc}
Although, as discussed above, there are some differences between the
canonical simulational estimates and our present Bethe lattice
calculations, they have many similarities. It is worth mentioning that
in the simulations no transitions were found for the RF model on the
{\em square} lattice \cite{kpor06}. It is possible that actually the
model shows a qualitatively different behavior on two-dimensional
lattices than the one found here, since the Bethe lattice may be
regarded as an infinite dimensional lattice \cite{b82}. 

Another
question which is worth to be considered is the relation of the model with
multiple monomers per site with the problem of the collapse transition
for polymers in a poor solvent. As mentioned above, one of the
simplest models used to study these transition is the SASAW's model,
so that it is interesting to find a relation between both models. We
may notice that the real situation of a polymer in a poor solvent may be
discretized by supposing the volume to be composed by cells of roughly
the size of a monomer, so that each cell will be either occupied by a
monomer (full) or by a solvent molecule (empty). For simplicity, we
are assuming the solvent molecules to have roughly the same size as
the monomers. Now if we require the
cells to form a regular lattice, we end up with a lattice gas
model. The monomer-monomer, monomer-solvent and solvent-solvent
interactions may then be considered effectively by introducing an
attractive interaction between monomers in first-neighbor sites which
are not connected by a polymer bond. Now we could imagine larger
cells, composed by $K$ of the original cells, so that each of them may
be occupied by up to $K$ monomers. If we now add the constraint that
no polymer bond may be formed between monomers in the same cell and
that attractive interactions only between monomers in the same cell
will be considered we end up essentially with the MMS model. The
parameters in the grand-canonical SASAW's models are the activity of a
monomer $z$ and the attractive interaction $-\epsilon$ ($\epsilon>0$)
between monomers. Considering the correspondence of this model with
the MMS model, we notice that the total contribution of a site with
$i$ monomers to the internal energy will be $-\epsilon \,i(i-1)/2$, so
that we may relate the parameters of both models as follows:
\begin{equation}
\omega_i=z^i\omega^{e(i)}, i=1,2,\ldots,K,
\end{equation}
where the exponent $e(i)=i(i-1)/2$ and
$\omega=exp(-\epsilon/k_BT)$. Therefore, we notice that the MMS model
with up to $K$ monomers per site corresponds to a grand-canonical
SASAW's model with constraints in a two-dimensional subspace of its
original $K$-dimensional parameter space. If we consider the canonical
situation, the dimensionality of the parameter space is reduced by one
in both models. 

In the particular case of the $K=3$ MMS model, we have the relations
$\omega_1=z$, $\omega_2=z^2\omega$, and $\omega_3=z^3\omega^3$. Thus,
recalling the definitions of the canonical parameters $\beta_i$ of the
model Eq. (\ref{beta}), it will correspond to the canonical SASAW's
model with constraints for $\beta_2=3\beta_1$. For the Bethe lattice
solution presented here, the multicritical point is located at
the origin in the $(\beta_1,\beta_2)$ space, and this point belongs to
the SASAW's subspace. However, as discussed above, this unphysical
localization of the multicritical point may be due to the approximate
character of the solution. In the simulations by Krawczyk et
al. of the RF model on the cubic lattice the
multicritical point is located in the first quadrant of the
$(\beta_1,\beta_2)$ space (Fig. 2 of reference
\cite{kpor06}). Unfortunately the precision in the estimated location
of the multicritical point in the simulations seems not to be
sufficient to determine its situation with respect to the
$\beta_2=3\beta_1$ line. It would be very interesting to find out if
the multicritical point is located {\em above} the SASAW's line, in
which case the collapse transition can be identified with the point
where the line crosses the tricritical line of the MMS model. In this
case, the collapse transition in the MMS model would be a tricritical
point, which is consistent with the well established result for this
transition.

\begin{figure}[h!]
\vspace{1.0cm}
\centering
\includegraphics[width=4cm]{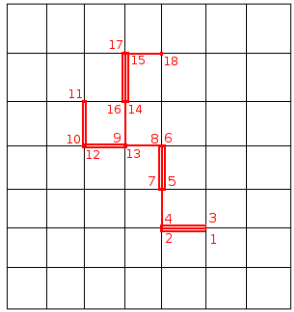}
\caption{(Color online) Contribution for the $K=2$ RA model on a square
  lattice, with a chainlike structure where all sites visited by the
  polymer are occupied by two monomers. The numbers correspond to the
  sequence of monomers in the chain.} 
\label{example}
\end{figure}

Finally, we notice that we have not studied the RA model here, where
immediate reversals are allowed. In the simulations of the $K=3$ case
of this model on the cubic lattice, no transition to the collapsed
phase was found \cite{kpor06}, suggesting that the RF constraint is
essential to produce this transition.  One possible explanation of the
reason for the effect 
of the RF constraint on the MMS model is that without this constraint
contributions are possible that actually correspond to extended
chainlike structures. Let us illustrate this by an example for the
$K=2$ model on a square lattice. If we have $\omega_2 \gg \omega_1$,
beside the {\bf DO} phase, where a pair of parallel bonds starts on the
surface and crosses the lattice, if immediate reversals are allowed
other chainlike structures, with a much higher entropy, are possible
with double occupied sites only, as may be seen in
Fig.\ref{example}. Such a contribution has an exponent $\nu$ which
correspond to extended polymers, and if these contributions dominate
in the polymerized phase no extended-collapsed transition will
occur. Of course, this argument is speculative and should be verified
by simulations or approximate calculations for the RA model.
As a final remark, we notice that no transition was found in the
simulations of the $K=3$ model on the square lattice
\cite{kpor06}. Although mean-field like calculations such as the one
presented here become less reliable as the dimension is lowered, thus
making it possible that e transition found in those approximations is
actually absent in the two-dimensional case, it is worth remaining
that the model of SASAW's on the square lattice is well studied and
shows a tricritical collapse transition \cite{ds85}, and therefore it
is interesting to further investigate the MMS model on two-dimensional
lattices. 

\section*{Acknowledgements}

TJO acknowledges doctoral grants by CNPq and FAPERJ
JFS acknowledges travel
support of the Argentinian agencies SECYTUNC and CONICET, and thanks
the Universidad Nacional de C\'ordoba for hospitality, he is also
grateful to CNPq for partial financial support. PS acknowledges
SECYT-UNC,  CONICET and FONCyT for partial financial support.

\appendix
\section{Coexistence surfaces in Bethe lattice calculations}
\label{ap}
The determination of the coexistence {\em loci} for solutions on
hierarchical lattices such as the Bethe and Husimi lattices presents
some difficulties, which may be related to the fact that in such
lattices one may readily obtain mean values in the central region, but
it is not straightforward to obtain the free energy (particularly the
entropy) as a mean value. One possibility is to integrate the state
equations to obtain the free energy, this may be even performed
analytically for some simple models such as the Ising model
\cite{b82}. In other cases it is possible to perform the integration
numerically, using a Maxwell construction to locate the discontinuous
transition. A detailed discussion of this point was presented by
Pretti \cite{p03}, analyzing particularly the proposals of Gujrati to
obtain the bulk free energy which was presented above \cite{g95} and
the one of Monroe, based on the Jacobian of the recursion relations
at the fixed point \cite{m94}. Also, a simple recursive criterion was
used to find the coexistence locus for a model of SASAW's on the
Husimi lattice, which consists in iterating the recursion relations
imposing `natural' initial conditions on the ratios. In the region
of the parameter space where the coexistence locus is located, there
are at least two stable fixed points, and the coexistence surface is
proposed to be the boundary of the basins of attraction of the fixed
points when the iteration is started with the `natural' initial
conditions. In the previous studies of the MMS model on Bethe and
Husimi lattices, the recursive procedure was used, but as shown above
it lead to results which are even qualitatively different from the
ones obtained using the more controlled approach by Gujrati. Here we
will discuss these questions for a simpler model than the one above:
SASAW's on the Bethe lattice, which was studied some time ago
\cite{ss90}.

The model is defined in the grand-canonical ensemble, so that $z$ will
be the activity of a monomer incorporated in the chains. The endpoints
of the chains are constrained to be on the surface of the tree. A
Boltzmann factor $\omega>1$ is associated to each pair of monomers in
first neighbor sites of the tree which are not connected by a polymer
bond, to take care of the attractive interactions. As usual, we define
partial partition functions for subtrees. The subtrees have a edge at
the root which is connected to the root site, to which $q-1$ subtrees
of the previous generation are attached. We call $g_0$
the partial partition functions for a subtree with no monomer on its
root site, $g_1$ will be the 
partial partition functions for a tree with a monomer on the root site
and no bond on 
the root edge, while $g_2$ stands for the partial partition functions
of a subtree with 
a monomer on the root site and a bond on the root edge. The recursion
relations for these partial partition functions are easily found to be:
\begin{subequations}
\begin{eqnarray}
g'_0\,&=&\, ( \,g_0 \,+\,g_1 \,)^{q-1} , \\
\mbox{}  \nonumber \\
g'_1\,&=&\,  \binom{q-1}{2} \, z \,g_2^2 \,(g_0+ \omega\,g_1)^{q-3}
\,, \\ 
\mbox{}   \nonumber \\
g'_2\,&=& \,(q-1) \,z  \, g_2 (g_0+ \omega\,g_1)^{q-2}  , 
\end{eqnarray}
\end{subequations}
Proceeding as usual, we may define the ratios $R_1=g_1/g_0$ and
$R_2=g_2/g_0$, and the recursion relations for them are:
\begin{subequations} 
\begin{eqnarray}
\label{rr1}
R'_1 \,&=& \,\binom{q-1}{2} \,z \,R_2^2 \, \frac{(1+
  w\,R_1)^{q-3}}{(1+R_1)^{q-1}}.\\
\mbox{} \nonumber \\
R'_2 \,&=&\,(q-1)  \, z \, R_2 \, \frac{(1 +
  w\,R_1)^{q-2}}{(1+R_1)^{q-1}},
\label{rr2}
\end{eqnarray}
\label{rr}
\end{subequations}
We should remark that the model was defined in a
different but equivalent way in the earlier calculation
\cite{ss90}, where an activity $x$ was associated to each {\em bond}
of the chains. Since all chains are long, as they are constrained to
start at the surface of the tree, these two formulations are
equivalent. For example, the recursion relations (3.8) in this 
reference correspond to the ones presented here if we relate the
ratios used in both calculations as $A=\sqrt{z}R_2$ and $B=R_1$, with
$x=(q-1)z$. 

The partition function of the model on the Cayley tree may be obtained
if we consider the operation of attaching $q$ subtrees to the central
site of the lattice:
\begin{equation}
Y\,=\, (g_0+g_1)^q+\binom{q}{2}\,z\,g_2^2\,(g_0+ \omega g_1)^{q-2},
\label{pfsasaw}
\end{equation}
and, following Gujrati's prescription Eq. (\ref{bfe}), 
the reduced bulk free energy per site is
\begin{equation}
\label{fesasaw}
\phi_b \,=\,
-\frac{1}{2}\left\{ q\ln{(1+R_1^*)^{q-1}}-(q-2)\,
\ln \left[ (1+R_1^*)^q+\binom{q}{2} z R_2^{* 2}
(1+\omega R_1^*)^{q-2} \right] \right\},
\end{equation}
\noindent where $(R_1^*,R_2^*)$ correspond to a fixed point values of 
the recursion relations Eqs.(\ref{rr}).

\subsection*{Non-polymerized fixed point}

In the  non-polymerized phase we have $R^{NP}_1\,=\,0$ and $R^{NP}_2\,=\,0$, 
the eigenvalues of the Jacobian are:

\begin{equation}
\lambda_1\,=\, \left. \frac{\partial\,R'_2}{\partial \,R_2}\right|_{NP} \,=\,
(q-1)\,z \;\;\;;\;\;\;
\lambda_2\,=\, \left. \frac{\partial\,R'_1}{\partial \,R_1}\right|_{NP} \,=\,0.
\end{equation}
Therefore, the stability limit of the non-polymerized fixed point will
be $z_{sl}^{NP}=1/(q-1)$. Using Eq.(\ref{fesasaw}) we see the free
energy vanishes 
for this phase, $\phi_b^{NP}=0$, as expected.

\subsection*{Polymerized fixed point}

In this phase $R_i^P \neq 0$, and $R_1^P$ can be obtained from Eq. (\ref{rr2}),
which in this case takes the form
\begin{equation}
\label{r1p}
(1+R_1^P)^{q-1} \,-\, (q-1)\,z\,(1+\omega \, R_1^P)^{q-2} \,=\,0 \,.
\end{equation}
For the particular case $q=3$ the fixed point equation above is
quadratic, and a simple expression may be found for the fixed point
value of $R_1$:
\begin{equation}
R_1^P=\omega z-1+\sqrt{(\omega z-1)^2-1+2z}.
\label{r1pol}
\end{equation}
The other root of the equation corresponds to an unstable fixed point.
Once $R_1^P$ is obtained, $R_2^P$ may be found using the other fixed
point equation, related to the recursion relation Eq. (\ref{rr1}), and
is given by 
\begin{equation}
\label{r2p}
(R_2^P)^2 \,=\,\frac{2}{q-2}\,R_1^P\,(1+\omega \, R_1^P)\, .
\end{equation}
The stability limit of this phase may be found by requiring the
largest eigenvalue of the Jacobian of the recursion relations
Eqs. \ref{rr} to be equal to one. In general, the equation defining
this limit of stability has to be solved numerically, but for $q=3$ it
is simple to find the result:
\begin{subequations}
\begin{eqnarray}
z_{SL}^P&=&\frac{1}{2},\;\; \mbox{for $\omega<2$,} \\
z_{SL}^P&=&\frac{2(\omega-1)}{\omega^2},\;\;\mbox{otherwise.}
\end{eqnarray}
\end{subequations}

The  tricritical point is obtained as the point on
the critical line $x=1/(q-1)$  where $R_1=0$ is a 
double root of  Eq.(\ref{r1p}):
\begin{equation}
z_{TC}\,=\,\frac{1}{q-1} \;\;\; \omega_{TC}\,=\,\frac{q-1}{q-2}
\end{equation}

\subsection*{The first-order line}

For $\omega > \omega_{TC}$ the first-order transition line can be
obtained using
the condition $\phi_b^P(x,\omega)\,=\,\phi_b^{NP}=0$, which gives
\begin{equation}
\label{fol}
(1\,+\,R_1^P)^{ 2\,(q-1)}\,=\,
\left[ 1\,+\,R_1^P\,+\,\frac{q}{q-2} \,
(1\,+\,\omega \,R_1^P) \,R_1^P \right]^{q-2},
\end{equation}
In general, this equation has to be  solved numerically, but for $q=3$
it is straightforward to obtain an analytical solution, which is:
\begin{equation}
R_1^{FO} \,=\, \sqrt{3 \, \omega -2}\,-\,2  \;,
\end{equation}
and, from Eq. (\ref{r1pol}) we obtain, 
\begin{equation}
\label{zfo}
z^{FO}(\omega) \,=\, \frac{(\sqrt{3 \, \omega -2}\,-\,1)^2}{2 (\omega \,
\sqrt{3 \, \omega -2}\,-\, 2\,\omega +1)}   \;\;\;;
\omega\,\ge \omega_{TC}\,=\,2.
\end{equation}

\begin{figure}[t]
\centering
\includegraphics[width=8cm]{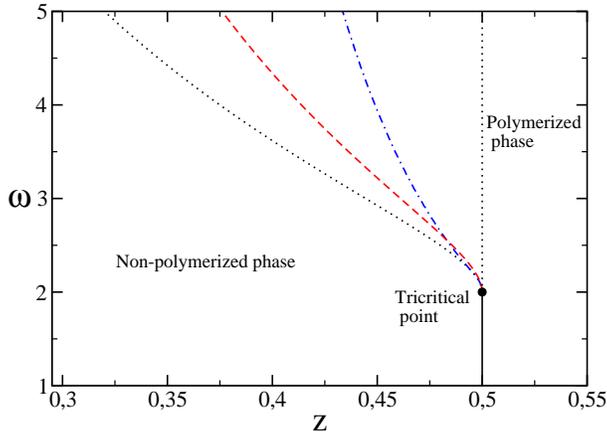}
\caption{(Color online) Phase diagram for SASAW's on a Bethe lattice
  with coordination $q=3$. Full (black) line is a continuous
  polymerization transition. Dotted lines are the limits of stability
  of the non-polymerized (right) and polymerized (left) phases. The
  dashed (red) line is the coexistence locus between both phases
  obtained from the free energy, while the dash-dotted (blue) line is
  the coexistence locus estimated using the recursive procedure.}  
\label{pdsasaws}
\end{figure}

The first order line calculated above does coincide with the one
obtained in the earlier calculation \cite{ss90} using the equal area
rule. This is expected, since the densities of monomers and of
interactions in the central region of the Cayley tree, calculated
directly from the partition function Eq. (\ref{pfsasaw}) are equal to
the ones obtained from the bulk free energy per site
Eq. (\ref{fesasaw}). Let us now consider the iterative prescription
suggested by Pretti in \cite{p02}. One iterates the recursion
relations Eqs. (\ref{rr}) starting with `natural' initial conditions for
the partial partition functions $g_0^{(0)}=1$, $g_1^{(0)}=0$, and
$g_2^{(0)}=z$, so that the initial values for the ratios are
$R_1^{(0)}=0$ and $R_2^{(0)}=z$. One then estimates the coexistence
line to be at the point in the $(z,\omega)$ parameter space where the
fixed point reached iterating the recursion relations switches between
the non-polymerized and polymerized phases. The results of all
calculations for a lattice with $q=3$ are displayed in the phase
diagram in Fig. \ref{pdsasaws}. The NIC method leads
to a first order line which is clearly different from the one obtained
using the bulk free energy. Actually, in more complex models, the
`natural' initial conditions may not be unique, and different choices
for them could lead to different results for the coexistence {\em
  locus}. The other methods to define the coexistence locus are
defined solely by the recursion relations and the partition
function. Also, despite the intuitive physical appeal of the method, 
its justification based on more solid arguments is still lacking.

Although in the SASAW's model discussed here the iterative procedure
has lead only to a quantitative error in the location of the first
order line, qualitative differences can result in more involved
models. An example is the MMS model. The {\bf DO} and {\bf TO} fixed
points, do not appear in the phase diagram despite the fact that the
fixed points associated to them are actually stable in regions of the
parameter space. This is due to the fact that the free energy of the
{\bf P1} and {\bf P2} phases is smaller in those regions. If, however,
the iterative procedure is used, this will be no longer the case and
those phases actually appear in the phase diagram, as may be seen in
the particular case $\omega_3=0$ in reference \cite{ss07}. For $K=3$
both {\bf DO} and {\bf TO} phases appear in the phase diagram if
the recursive procedure is used.

Finally, we will briefly discuss the suggestion by Monroe \cite{m94}
that at the  
coexistence the leading eigenvalues of the Jacobian of both phases
should be equal. We may consider points on the coexistence line
Eq. (\ref{zfo}) for $q=3$. If the eigenvalue which corresponds to the
polymerized phase would be equal do the one associated to the
non-polymerized phase $\lambda_{NP}=2z$, then the expression:
\begin{equation}
\Delta=(J_{1,1}-2z)(J_{22}-2z)-J_{1,2}J_{2,1},
\end{equation}
where $J_{i,j}$ are the elements of the Jacobian calculated at the
polymerized fixed point, should vanish on the coexistence line. It may
be shown that this expression does not vanish for $\omega>2$, thus
showing that the Monroe criterium is not equivalent to the free-energy
criterium for this particular model. It is interesting to remark that
in other models this equivalence was found \cite{p03}.

\section{Location of the multicritical point}
\label{lmp}
To find the location of the multicritical point in the parameter space
of the model, we look for higher order roots with vanishing ratios of
the fixed point equations which follow if we set $R'_i=R_i$ in the
recursion relations Eqs. \ref{rrrf}. An inspection of the fixed point
equations suggests the following ansatz for the ratios close to the
non-polymerized fixed point: $R_2=aR_1^2$ and $R_3=bR_1^3$. We then
substitute these leading order terms into the fixed point equations and
require them to be satisfied up to order 4 in $R_1$. This furnishes
five equations: from the first recursion relation Eq. (\ref{rrr1}) we
get one equation for order $R_1$ and another for order $R_1^2$. The
second recursion relation Eq. (\ref{rrr2}) furnishes two equations, one
for order $R_1^2$ and the other for order $R_1^4$. Finally, recursion
relation Eq. (\ref{rrr3}) provides an additional equation for order
$R_1^3$. In the sequence adopted above, the equations are:
\begin{subequations}
\begin{eqnarray}
1 &=& (q-1)\omega_1, \\
\binom{q-1}{2} \, \omega_1 &=& 3\,\binom{q-1}{3}\,\omega_2+4\,
\binom{q-1}{2} \, \omega_2 a, \\
a &=& \binom{q-1}{2}\,\omega_2+(q-1)\omega_2a, \\
\binom{q-1}{2}\,\omega_1a &=& 6\,\binom{q-1}{4}\,\omega_3+15 \,
\binom{q-1}{3}\, \omega_3a+4\,\binom{q-1}{2}\,\omega_3a^2 + \nonumber
\\ 
&&6\,\binom{q-1}{2}\, \omega_3 b, \\
b &=& \binom{q-1}{3}\, \omega_3+2\, \binom{q-1}{2}\, \omega_3a + (q-1)
\omega_3b.
\end{eqnarray}
\end{subequations}
These equations may easily be solved, leading to $\omega_1=1/(q-1)$,
$\omega_2=1/(q-1)^2$, $\omega_3=1/(q-1)^3$, $a=1/2$, and $b=1/6$. The
behavior of the ratios in the vicinity of the multicritical point has
been verified numerically.


\begin{thebibliography}{99}
\bibitem{f66}P. J. Flory, {\em Principles of Polymer Chemistry},
  $5^{th}$ edition, Cornell University Press, NY (1966).
\bibitem{dg72}P. G. de Gennes, Phys. Lett. A {\bf 38},339 (1972).
\bibitem{dg79}P. G. de Gennes, {\em Scaling Concepts in Polymer
    Physics}, Cornell University Press, NY (1979).
\bibitem{dg75}P. G. de Gennes, J. Physique Lettres {\bf 36}, 1049
  (1975). 
\bibitem{ds85}B. Derrida and H. Saleur, J. Phys. A {\bf 18}, L1075
  (1985); H. Saleur, J. Stat. Phys. {\bf 45}, 419 (1986).
\bibitem{ds87}B. Duplantier and H. Saleur, Phys. Rev. Lett {\bf 59},
  539 (1987); B. Duplantier, Phys. Rev. A {\bf 38}, 3647 (1988).
\bibitem{sms96}J. F. Stilck, K. D. Machado, and P. Serra,
  Phys. Rev. Lett. {\bf 76}, 2734 (1996); J. F. Stilck, P. Serra, and  
  K. D. Machado, Phys. Rev. Lett. {\bf 89}, 169602 (2002);
  P. Serra, J. F. Stilck, W. L. Cavalcanti, and K. D. Machado,
  J. Phys. A {\bf 37}, 8811 (2004);K. D. Machado, M. J. de Oliveira,
  and J. F. Stilck, Phys. Rev. E {\bf 64}, 051810 (2001); D. P. Foster,
  J. Phys A {\bf 40}, 1963 (2007).
\bibitem{p02}M. Pretti, Phys. Rev. Lett. {\bf 89}, 169601 (2002).
\bibitem{kpor06}J. Krawczyk, T. Prellberg, A. L. Owczarek, and
  A. Rechnitzer, Phys Rev. Lett. {\bf96}, 240603 (2006).
\bibitem{dj72}C. Domb and G. S. Joyce, J. Phys. C {\bf 5}, 956
  (1972). 
\bibitem{ss07}P. Serra and J. F. Stilck, Phys. Rev. E {\bf 75}, 
  011130 (2007).
\bibitem{oss08}T. J. Oliveira, J. F. Stilck, and P. Serra
  Phys. Rev. E {\bf 77}, 041103 (2008).
\bibitem{g95}P. D. Gujrati, Phys. Rev. Lett. {\bf 74}, 809 (1995).
\bibitem{gw70}R. B. Griffiths and J. C. Wheeler, Phys. Rev. A {\bf 2},
  1047 (1970); R. B. Griffiths, Phys. Rev. B{\bf 7}, 545 (1973).
\bibitem{ls84}I. D. Lawrie and S. Sarbach in {\em Phase Transitions
  and Critical Phenomena}, vol. {\bf 9},  ed. by C. Domb and
  J. L. Lebowitz, Academic Press (1984).
\bibitem{g00}D. MacDonald et al, J. Phys. A {\bf 33}, 5973 (2000).
\bibitem{b82}R. J. Baxter, {\em Exactly Solved Models in 
  Statistical Mechanics}, Academic, London (1982).
\bibitem{p03}M. Pretti, J. Stat. Phys. {\bf 111}, 993 (2003).
\bibitem{m94}J. L. Monroe, Phys. Lett A {\bf 188}, 80 (1994).
\bibitem{ss90}P. Serra and J. F. Stilck, J. Phys. A {\bf23}, 5351
  (1990). 

\end{thebibliography}
\end{document}